\documentclass
[superscriptaddress,secnumarabic,amssymb,amsmath,nobibnotes,aps,prd,showkeys,showpacs,nofootinbib]{revtex4}%
\usepackage{graphicx}
\usepackage{epsf}
\usepackage{bm}
\usepackage{amsmath}
\usepackage{amsfonts}
\usepackage{amssymb}
\usepackage{epstopdf}%
\providecommand{\U}[1]{\protect\rule{.1in}{.1in}}
\newcommand{\be}{\begin{equation}}
\newcommand{\ee}{\end{equation}}

\newcommand{\mincir}{\raise
-3.truept\hbox{\rlap{\hbox{$\sim$}}\raise4.truept\hbox{$<$}\ }}
\newcommand{\magcir}{\raise
-3.truept\hbox{\rlap{\hbox{$\sim$}}\raise4.truept\hbox{$>$}\ }}

\begin{document}
\title{Exact solution of the Einstein-Skyrme model in a Kantowski-Sachs spacetime}
\author{Andronikos Paliathanasis}
\email{anpaliat@phys.uoa.gr}
\affiliation{Instituto de Ciencias F\'{\i}sicas y Matem\'{a}ticas, Universidad Austral de
Chile, Valdivia, Chile}
\author{Michael Tsamparlis}
\email{mtsampa@phys.uoa.gr}
\affiliation{Faculty of Physics, Department of Astronomy-Astrophysics-Mechanics University
of Athens, Panepistemiopolis, Athens 157 83, Greece}

\begin{abstract}
We consider a Skyrme fluid with a constant radial profile in locally
rotational Kantowski-Sachs spacetime. The Skyrme fluid is an anisotropic fluid
with zero heat flux and with an equation of state parameter $w_{S}$ that
$\left\vert w_{s}\right\vert \leq\frac{1}{3}$. From the Einstein field
equations we define the Wheeler-DeWitt equation. For the last equation we
perform a Lie symmetry classification and we determine the invariant solutions
for the wavefunction of the model. Moreover from the Lie symmetries of the
Wheeler-DeWitt equation we construct Noetherian conservation laws for the
field equations which we use in order to write the solution in closed form. We
show that all pf the cosmological parameters are expressed in terms of the
scale factor of the two dimensional sphere of the Kantowski-Sachs spacetime.
Finally from the application of Noether's theorem for the Wheeler-DeWitt
equation we derive conservation laws for the wavefunction of the universe.

\end{abstract}
\maketitle

\section{Introduction}

\label{intro}

The Skyrme model has various applications in physics (for instance see
\cite{ref1,ref2,ref3,ref4,ref5}). The model which was proposed by Skyrme in
the early of 60s \cite{Skyrme1,Skyrme2} is a nonlinear theory of pions in
which the baryons can be interpreted as topological soliton solutions of the
model in the configuration space \cite{TopSol}. The Skyrme field has been used
as a sourced fluid in General Relativity, the so-called Einstein-Skyrme model.
It has been shown that the Einstein-Skyrme model admits static black hole
solutions with a regular event horizon which approaches asymptotically the
Schwarzschild solution; furthermore, the Einstein-Skyrme model violates the
\textquotedblleft no hair\textquotedblright\ conjecture for black holes
\cite{bh01,bh02,bh03,Ioannidou}.

Recently the study of the dynamical evolution of the field equations of the
Einstein-Skyrme model with a cosmological constant in a locally rotational
\ Kantowski-Sachs spacetime was performed in \cite{Parisi}. In particular for
the Skyrme fluid a constant radial profile oon the hedgehog Ansatz has been
considered\footnote{For a discussion of the hedgehod ansatz on spherically
symmetric spacetime and some generalizations see
\cite{Canfora,Canfora2,Canfora3}.} and it has been shown that the field
equations admit two stable point solutions. The two solutions describe an
exponential scale factor for the

two-dimensional sphere of the Kantowski-Sachs spacetime. For a nonconstant
radial profile, numerical solutions of the Einstein-Skyrme model in
anisotropic cosmologies have been studied in \cite{Canfora4}. Specifically, in
\cite{Canfora4}, the Einstein-Skyrme model was studied in Bianchi I and in
Kantowski-Sachs universes and it has been found that bounds on the values of
the cosmological constant and on the Skyrme coupling should be taken in order
for solutions to exist.

In this work we consider the Einstein-Skyrme model with a constant radial
profile and show that the Wheeler-DeWitt (WdW) Equation can be solved
explicitly by means of the use of Lie point symmetries. Furthermore, following
the method which was established in \cite{AB} for scalar field cosmology with
a matter source, we determine Noetherian conservation laws for the field
equations. The existence of the new conservation laws implies that the field
equations are Liouville integrable and with the method of separation of
variables we write the analytical solution of the model.

Lie and Noether point symmetries are powerful tools which have been used for
the study of various models in gravitational physics and cosmology. The
symmetry analysis of charged perfect fluids in spherically symmetric
spacetimes can be found in \cite{Maharaj1,Maharaj3,MahLeach}. On the other
hand Noether's Theorem has been applied in various cosmological models in
order to constrain the unknown functions in scalar-field cosmology, in
$f\left(  R\right)  $ gravity, in $f\left(  T\right)  $ gravity and others (
see \cite{deRiti,KotsakisL,VakF,dimakisT,dimakisT2,Dib,Hwei,PalFR} and
references therein). The plan of the paper is as follows.

In Section \ref{field} we consider our model which is General Relativity with
a cosmological constant in a locally rotational Kantowski-Sachs spacetime
where the matter source is a Skyrme fluid. For that model we write the field
equations and we derive the WdW Equation. The existence of Lie point
symmetries for the WdW equation and the invariant solutions are given in
Section \ref{wdweq}. In Section \ref{classicalSolution} we follow the method
which was established in \cite{AB} and we use the Lie point symmetries of the
WdW Equation in order to study the Liouville integrability of the field
equations. We construct the closed-form classical solution of the model, and
we show that the cosmological parameters can be written in terms of the scale
factor of the two-dimensional sphere, while the quantum potential is
calculated. Finally, in Section \ref{Conc} we draw our conclusions.

\section{Einstein-Skyrme Action}

\label{field}

Consider the Einstein-Skyrme Action
\begin{equation}
S=\int dx^{4}\sqrt{-g}\left(  R-2\Lambda\right)  +S_{K}, \label{eq.00}%
\end{equation}
in a Kantowski-Sachs background with line element%
\begin{equation}
ds^{2}=-dt^{2}+A^{2}\left(  t\right)  dr^{2}+B^{2}\left(  t\right)  \left(
d\theta^{2}+\sin^{2}\theta d\phi^{2}\right)  . \label{eq.0}%
\end{equation}
Spacetime (\ref{eq.0}) is a locally rotational spacetime (LRS) and admits a
four-dimensional Killing algebra. In \ particular, it admits as Killing
vectors the Lie group $SO\left(  3\right)  $ plus the Killing vector
$\partial_{r}$.

The term, $S_{K},$ in the total action (\ref{eq.00}) is the action of the
Skyrme field%
\begin{equation}
S_{K}=\frac{K}{2}\int dx^{4}\sqrt{-g}\left(  \frac{1}{2}R_{\mu}R^{\mu}%
+\frac{\lambda}{16}F_{\mu\nu}F^{\mu\nu}\right)  ,
\end{equation}
where $R_{\mu}=U^{-1}U_{;\mu}$,~$F_{\mu\nu}=\left[  R_{\mu},R_{\nu}\right]  $
and $U\left(  x^{\mu}\right)  $ is the $SU\left(  2\right)  $-valued scalar;
finally $K~$and$~\lambda$ are positive constants~\cite{Canfora,TopSol}.

Variation with respect to the metric in (\ref{eq.00}) provides us with the
gravitational field equations%
\begin{equation}
G_{\mu\nu}+\Lambda g_{\mu\nu}=kT_{\mu\nu}, \label{eq.A}%
\end{equation}
in which $G_{\mu\nu}$ is the Einstein tensor $k$ is the Einstein constant, and
$T_{\mu\nu}$ is the energy-momentum tensor for the Skyrme fluid. Furthermore
variation with respect to the field $U\left(  x^{\mu}\right)  $ in
(\ref{eq.00}) provides the matter conservation law $T_{~;\nu}^{\mu\nu}=0$
which is%
\begin{equation}
g^{\mu\nu}\left(  R_{\mu;\nu}+\frac{\lambda}{4}\left(  \left[  R^{\sigma
},F_{\sigma\nu}\right]  \right)  _{;\nu}\right)  =0. \label{eq.00a}%
\end{equation}

\ Following \cite{Canfora,Parisi} we consider the hedgehog Ansatz in which the
radial profile function is constant and, specifically, $\frac{\pi}{2}%
+\sigma\pi,\sigma\in%
\mathbb{Z}
,$ where equation (\ref{eq.00a}) is satisfied indentically, and the remaining
system comprises the gravitational field equations (\ref{eq.A}) and now the
energy-momentum tensor, $T_{\mu\nu},$ has the following nonzero components
\cite{Canfora,Parisi}%
\begin{equation}
T_{\theta}^{\theta}=T_{\phi}^{\phi}=\frac{K\lambda}{2B^{4}} \label{TE.3a}%
\end{equation}
and%
\begin{equation}
T_{t}^{t}=T_{r}^{r}=-\frac{K}{B^{2}}\left(  1+\frac{\lambda}{2B^{2}}\right)  .
\label{TE.3b}%
\end{equation}

We recall that in the 1+3 analysis defined by the four-velocity $u^{a}$ the
energy-momentum tensor is decomposed as follows%
\[
T_{\mu\nu}=\rho_{S}u_{\mu}u_{\nu}+p_{s}h_{\mu\nu}+q_{(\mu}u_{\nu)}+\pi_{\mu
\nu},
\]
where
\begin{align}
\rho_{S}  &  =T_{\mu\nu}u^{\mu}u^{\nu}~,~p_{S}=\frac{1}{3}h^{\kappa\lambda
}T_{\kappa\lambda},\label{TE.1}\\
q^{\mu}  &  =h^{\mu\kappa}T_{\kappa\nu}u^{\nu}~,~\pi_{\mu\nu}=(h_{\mu}%
^{\kappa}h_{\nu}^{\lambda}-\frac{1}{3}h_{\mu\nu}h^{\kappa\lambda}%
)T_{\kappa\lambda} \label{TE.2}%
\end{align}
and
\begin{equation}
h_{\mu\nu}=g_{\mu\nu}+u_{\mu}u_{\nu}%
\end{equation}
is the $u^{a}$ projective tensor. The variables, $\rho_{S},p_{S},q^{\mu}%
,\pi_{\mu\nu},$ are the dynamical variables of the

energy-momentum tensor and represent respectively the mass density, the
isotropic pressure, the heat flux and the traceless stress tensor, as measured
by the observer $u^{\mu}$. Using this general result from (\ref{TE.3a}),
(\ref{TE.3b}) we have that $q^{\mu}=0$,
\begin{equation}
\rho_{S}=\frac{1}{B^{2}}\left(  \bar{k}+\frac{\mu}{2B^{2}}\right)
~,~p_{S}=-\frac{1}{3B^{2}}\left(  \bar{k}-\frac{\mu}{2B^{2}}\right)
\label{TE.3}%
\end{equation}
and%
\begin{equation}
\pi_{\mu\nu}=diag\left(  0,-\frac{2}{3}\rho_{S},\frac{1}{3}\rho_{S},\frac
{1}{3}\rho_{S}\right)  , \label{TE.4}%
\end{equation}
where the new constants $\bar{k}~$and$~\mu~$are given by the relations
$\bar{k}=kK$ and $\mu=\bar{k}\lambda.$

We define the equation-of-state parameter, $w_{S},$ for the Skyrme fluid as
follows%
\begin{equation}
w_{S}=\frac{p_{S}}{\rho_{S}}=-\frac{1}{3}\frac{\bar{k}B^{2}-\mu}{\bar{k}%
B^{2}+\mu}. \label{TE.5}%
\end{equation}
We note that, when $\bar{k}B>\mu$ and $w_{S}\simeq-\frac{1}{3},$ the fluid
acts as a curvature-like component, whereas for $\bar{k}B^{2}<<\mu$ and
$w_{S}\simeq\frac{1}{3}$ and the Skyrme fluid has the equation-of-state
parameter of a radiation-like fluid. It follows that the equation-of-state
parameter for the Skyrme fluid is bounded as follows%
\begin{equation}
-\frac{1}{3}\leq w_{S}\leq\frac{1}{3}.
\end{equation}

\subsection{Minisuperspace approach}

From the Action-integral, (\ref{eq.00}), and for the spacetime with line
element, (\ref{eq.0}), we find that the Lagrangian of the field equations is
given by the following expression%
\begin{equation}
L\left(  A,B,\dot{A},\dot{B}\right)  =2B\dot{A}\dot{B}+A\dot{B}^{2}%
+AB^{2}\Lambda+A\left(  \bar{k}-1\right)  +\frac{A\bar{k}\lambda}{2B^{2}}
\label{eq.01}%
\end{equation}
and that the field equations, (\ref{eq.A}), are \cite{Parisi}:%
\begin{equation}
2\frac{\dot{A}\dot{B}}{AB}+\frac{\dot{B}^{2}}{B^{2}}+\frac{1}{B^{2}}%
-\Lambda=\frac{1}{B^{2}}\left(  \bar{k}+\frac{\mu}{2B^{2}}\right)  ,
\label{eq.1}%
\end{equation}%
\begin{equation}
2\frac{\ddot{B}}{B}+\frac{\dot{B}^{2}}{B^{2}}+\frac{1}{B^{2}}-\Lambda=\frac
{1}{B^{2}}\left(  \bar{k}+\frac{\mu}{2B^{2}}\right)  \label{eq.2}%
\end{equation}

and%

\begin{equation}
\frac{\ddot{A}}{A}+\frac{\ddot{B}}{B}+\frac{\dot{A}\dot{B}}{AB}-\Lambda
=-\frac{\mu}{2B^{4}}. \label{eq.3}%
\end{equation}

Moreover, for the lapse time $dt=N\left(  \tau\right)  d\tau$, the Lagrangian
of the field equations (\ref{eq.01}) becomes%
\begin{equation}
L\left(  N,A,B,\dot{A},\dot{B}\right)  =\frac{1}{N}\left(  2BA^{\prime
}B^{\prime}+AB^{\prime2}\right)  +N\left(  AB^{2}\Lambda+\omega A+\mu\frac
{A}{2B^{2}}\right)  , \label{eq.5}%
\end{equation}
where now the new constant $\omega$ is $\omega=\bar{k}-1.$ In this a case the
field equations can be seen as the Euler-Lagrange equations with respect to
the variables $\left\{  N,A,B\right\}  ~$\cite{Ray}.

In the minisuperspace approach the WdW Equation is defined as follows
\cite{Will,Hall},%
\begin{equation}
\hat{H}\Psi=0, \label{eq.05a}%
\end{equation}
where $\hat{H}$ is the Hamiltonian operator defined by the conformal Laplace operator.

From the Lagrangian (\ref{eq.5}) we can define the momenta, $p_{A}$and
$p_{B},$ as%
\begin{equation}
p_{A}=\frac{2}{N}BB^{\prime}~~,~~p_{B}=\frac{2}{N}\left(  BA^{\prime
}+AB^{\prime}\right)  . \label{eq.6}%
\end{equation}

Hence the Hamiltonian function is%
\begin{equation}
H\equiv N\left[  -\frac{A}{4B^{2}}p_{A}^{2}+\frac{1}{2B}p_{A}p_{B}-\left(
AB^{2}\Lambda+\omega A+\mu\frac{A}{2B^{2}}\right)  \right]  =0. \label{eq.7}%
\end{equation}

Therefore under normal quantization, $p_{i}\sim\frac{\partial}{\partial x^{i}%
}$, from the Hamiltonian (\ref{eq.7}) and equation (\ref{eq.05a}) we derive
the WdW Equation\footnote{Recall that for the field equations (\ref{eq.1}%
)-(\ref{eq.3}) the minisuperspace has dimension two.}
\begin{equation}
W\left(  A,B,\Psi,\Psi_{B},\Psi_{,AA},\Psi_{,BB},\Psi_{,AB}\right)  \equiv0,
\label{eq.08}%
\end{equation}
where $W$ is%
\begin{equation}
W\equiv-\frac{A}{4B^{2}}\Psi_{,AA}+\frac{1}{2B}\Psi_{,AB}-\frac{1}{4B^{2}}%
\Psi_{,A}-\left(  AB^{2}\Lambda+\omega A+\mu\frac{A}{2B^{2}}\right)
\Psi\label{eq.09}%
\end{equation}
and $\Psi=\Psi\left(  A,B\right)  $ is the wavefunction of the universe. We
observe that as the dimension of the minisuperspace is two in the WdW
Equation, (\ref{eq.08}), there is no quantum correction term, i.e., the Ricci
Scalar of the minisuperspace is an extra potential term.~\ Furthermore we
remark that only the geometric degrees of freedom, i.e., $\left\{
A,B\right\}  $ are quantized and not the $SU\left(  2\right)  $ field from the
Skyrme-fluid. This is because the Skyrme-fluid \ is expressed on terms of the
parameters, $\left\{  A,B\right\}  $, and there is not another degree of
freedom which has been introduced from the Skyrme-fluid as we have considered
that the radiation profile in the hedgehog Ansatz is constant.

In order to solve equation (\ref{eq.09}) we apply the method of group
invariant transformations, specifically the Lie point symmetries. Recently in
\cite{AB} it has been shown that the existence of Lie point symmetries for the
WdW Equation is related with oscillatory terms in the wave function of the
universe. Furthermore, the existence of Lie point symmetries means that the
classical field equations admit Noetherian conservation laws which can lead to
the integrability of the field equation. \ In the following Section we proceed
with the determination of the Lie point symmetries for the WdW Equation,
(\ref{eq.08}), and with the application of Lie invariants in order to
construct analytical solutions of the wavefunction $\Psi\left(  A,B\right)  $.
Analytical solutions of equation (\ref{eq.08}) without the Skyrme term can be
found in \cite{Conradi,Lopez}.

\section{Point symmetries and invariant solutions for the WdW equation}

\label{wdweq}

The WdW Equation (\ref{eq.08}) is a second-order partial differential equation
with independent variables $\left\{  A,B\right\}  $ and dependent variable
$\Psi$. By definition the generator%
\begin{equation}
\mathbf{X}=\xi^{A}\left(  A,B,\Psi\right)  \partial_{A}+\xi^{B}\left(
A,B,\Psi\right)  \partial_{B}+\eta\left(  A,B,\Psi\right)  \partial_{\Psi},
\label{eq.10}%
\end{equation}
of a one-parameter point transformation on the space of variables $\left\{
A,B,\Psi\right\}  $ is called a Lie point symmetry of the differential
equation $W$ when there exist a function $\lambda$ such that the following
condition holds \cite{Bluman}%
\begin{equation}
\mathbf{X}^{\left[  2\right]  }W=\lambda W,~ \label{eq.11}%
\end{equation}
where $X^{\left[  2\right]  }$ is the second prolongation/extension of
$\mathbf{X}$ in the space $\left\{  A,B,\Psi,\Psi_{,A},\Psi_{,B}\right\}  $.

The symmetry condition, (\ref{eq.11}) provides a polynomial system on the
derivatives of $\Psi$ the solution of which gives the unknown functions
$\xi^{A},\xi^{B}$ and $\eta.~$Moreover, because\ (\ref{eq.08}) is a linear
second-order partial differential equation of dimension two, the field
(\ref{eq.10}) has the following form%
\begin{equation}
\mathbf{X}=\mathbf{\xi}+\left(  a_{0}\Psi+b\left(  A,B\right)  \right)
\partial_{\Psi}, \label{eq.12}%
\end{equation}
where $b\left(  A,B\right)  $ is a solution of the original equation and
$\mathbf{\xi}=\xi^{A}\left(  A,B\right)  \partial_{A}+\xi^{B}\left(
A,B\right)  \partial_{B}.$

From condition (\ref{eq.11}) for the WdW Equation (\ref{eq.08}) we find the
following Lie point symmetries,%
\begin{equation}
X_{\Psi}=\Psi\partial_{\Psi}~,~X_{b}=b\partial_{\Psi}~,~X_{1}=\frac{1}%
{AB}\partial_{A}, \label{eq.13}%
\end{equation}%
\begin{equation}
X_{2}=\frac{B}{4\Lambda B^{4}+4\omega B^{2}+2\mu}\left(  -A\partial
_{A}+2B\partial_{B}\right)  \label{eq.14}%
\end{equation}
and%

\begin{equation}
X_{3}=\left(  \frac{2AB^{2}\left(  3\omega+2\Lambda B^{2}\right)  }{3\left(
2\Lambda B^{4}+2\omega B^{2}+\mu\right)  }\partial_{A}-\frac{\left(  2\Lambda
B^{4}+6\omega B^{2}-3\mu\right)  B}{3\left(  2\Lambda B^{4}+2\omega B^{2}%
+\mu\right)  }\partial_{B}\right)  , \label{eq.15}%
\end{equation}
with nonzero Lie Brackets~$\left[  X_{1},X_{3}\right]  =X_{1},~\left[
X_{2},X_{3}\right]  =-X_{2}$. The vector fields, $X_{\Psi},X_{b},$ are trivial
symmetries and reflect that the differential equation, $W,$ is
linear\footnote{The Lie point symmetry $X_{\Psi}$ is called linear symmetry,
and $X_{b}$ express the infinity number of solutions.}.

The existence of a (nontrivial) Lie symmetry vector for the second-order
partial differential equation, $W$, means that there exists a coordinate
transformation, $\left\{  A,B\right\}  \rightarrow\left\{  \bar{A},\bar
{B}\right\}  $, in which equation $W$ is independent of one of the independent
variables $\left\{  \bar{A},\bar{B}\right\}  ,~$i.e.,~$\frac{\partial
W}{\partial\bar{A}}=0$~or~$\frac{\partial W}{\partial\bar{B}}=0$, or
equivalently, that there exists a coordinate transformation $\left\{
A,B,\Psi\right\}  \rightarrow\left\{  \bar{A},\bar{B},\bar{\Psi}\right\}  ,$
in which the partial differential equation, $W$, reduces to an ordinary
differential equation\footnote{In general a partial differential equation with
$l-$dependent variables is reduced to a differential equation of $\left(
l-1\right)  $-dependent variables.}. Solutions which follow from the
application of the Lie point symmetries are called invariant solutions.

The application of the Lie point symmetry, $Y_{1}=X_{1}+m_{1}X_{\Psi}%
,~m_{1}\in%
\mathbb{C}
,$ to (\ref{eq.08}) gives that%
\begin{equation}
\Psi_{1}\left(  A,B\right)  =\exp\left(  \frac{m_{1}}{2}A^{2}B\right)
\Phi\left(  B\right)  ,
\end{equation}
where $\Phi\left(  B\right)  $ satisfies the following ordinary differential
equation%
\begin{equation}
m_{1}B^{2}\Phi_{,B}-\left(  2B^{4}\Lambda+2\omega B^{2}+\mu\right)  =0.
\end{equation}
Hence we find that the invariant solution which follows from the application
of the symmetry vector $Y_{1}$ is%
\begin{equation}
\Psi_{1}\left(  A,B\right)  =\Psi_{1}^{0}\exp\left(  \frac{2\Lambda
B^{4}+6\omega B^{2}-3\mu}{3m_{1}B}+\frac{1}{2}m_{1}A^{2}B\right)
\label{eq.16}%
\end{equation}
for $m_{1}\neq0,~$and~$\Psi_{1}\left(  A,B\right)  =0$, for $m_{1}=0$.

Similarly, from the symmetry vector, $Y_{2}=X_{1}+m_{2}X_{\Psi}$,~$m_{2}\in%
\mathbb{C}
,$ we find that the invariant solution is of the form%
\begin{equation}
\Psi_{2}\left(  A,B\right)  =\exp\left(  -\frac{m_{2}}{6B}\left(  2\Lambda
B^{4}+6\omega B^{2}-3\mu\right)  \right)  \bar{\Phi}\left(  C\right)
~,~C=A^{2}B, \label{eq.16b}%
\end{equation}
where $\Phi\left(  C\right)  $ is given by the following equation%
\begin{equation}
m_{2}\bar{\Phi}_{,C}+\bar{\Phi}=0.
\end{equation}
Hence the invariant solution is,%
\begin{equation}
\Psi_{2}\left(  A,B\right)  =\Psi_{2}^{0}\exp\left(  -\frac{m_{2}}{6B}\left(
2\Lambda B^{4}+6\omega B^{2}-3\mu\right)  \right)  \exp\left(  -\frac{A^{2}%
B}{m_{2}}\right)
\end{equation}
for $m_{2}\neq0$, and $\Psi_{2}\left(  A,B\right)  =0$, for $m_{2}=0$.

\begin{figure}[ptb]
\includegraphics[height=10cm]{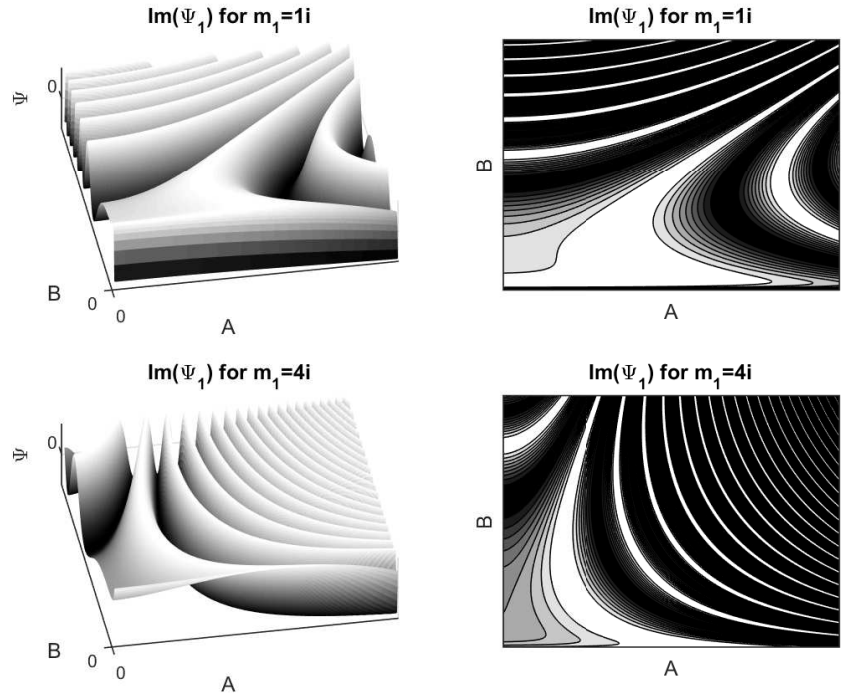}
\caption{Evolution of the invariant solution $\operatorname{Im}\left(
\Psi_{1}\right)  ~$of (\ref{eq.16}) which corresponds to the Lie point
symmetry $Y_{1}$, for $m_{1}=1i$, and, $m_{1}=4i$. The plots are in the plane
$A,B$, for $\Lambda=1$, $\omega=-0.5$ and $\mu=0.2$.}%
\label{fig2a}%
\end{figure}

\begin{figure}[ptb]
\includegraphics[height=10cm]{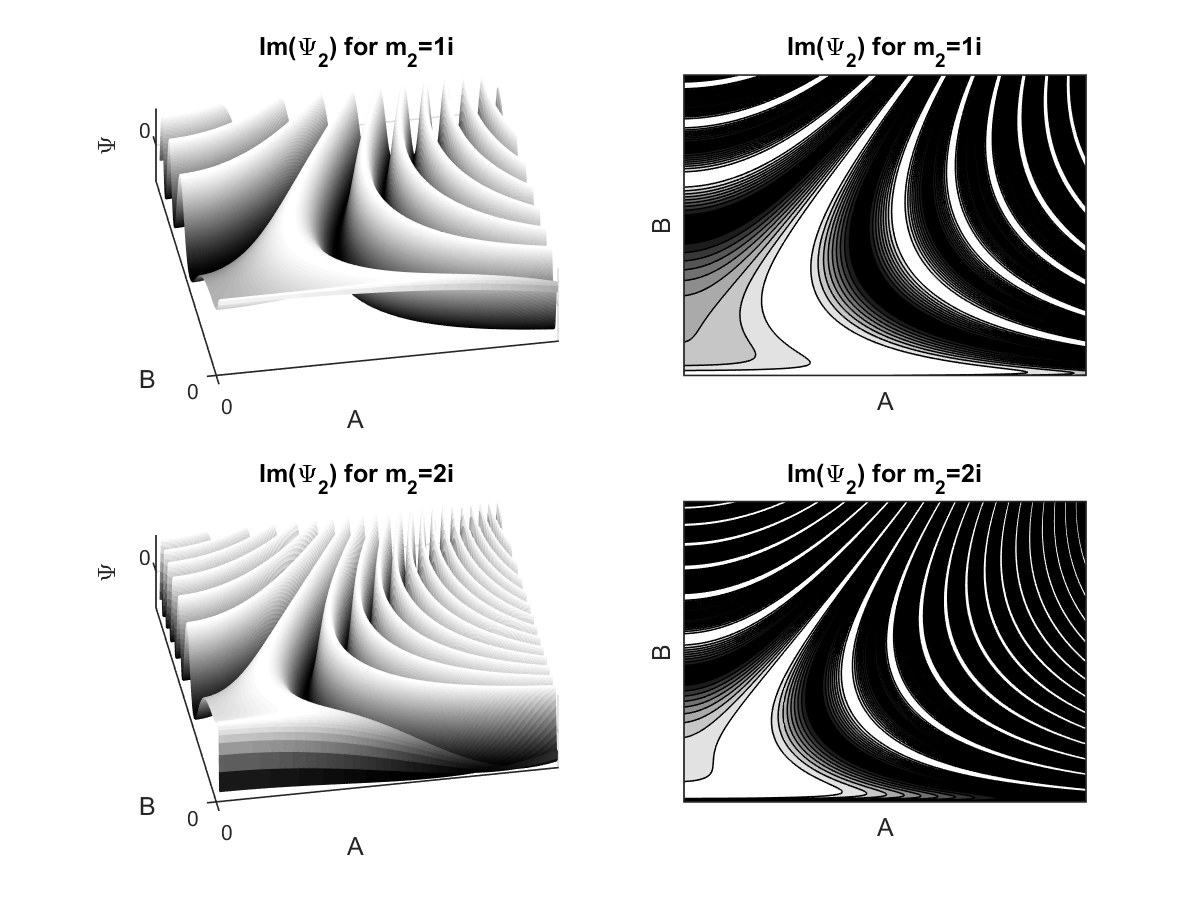}
\caption{Evolution of the invariant solution $\operatorname{Im}\left(
\Psi_{2}\right)  ~$of (\ref{eq.16b}) which corresponds to the Lie point
symmetry $Y_{2}$, for $m_{2}=1i$, and, $m_{2}=2i$. The plots are in the plane
$A,B$, for $\Lambda=1$, $\omega=-0.5$ and $\mu=0.2$.}%
\label{fig2b}%
\end{figure}

In Figs. \ref{fig2a} and \ref{fig2b} the qualitative evolution of the
imaginary part of the wavefunction for the invariant solutions, $\Psi
_{1}\left(  A,B\right)  $~and $\Psi_{2}\left(  A,B\right)  $, are given
respectively for two different values of the constants $m_{1}~$and $m_{2}$.
For the figures we considered the values $\Lambda=1$, $\omega=-0.5$ and
$\mu=0.2$. From the figures the oscillatory behaviour of the wavefunction,
which is related to the existence of Lie symmetries, can be observed.


From the Lie symmetry vector, $Y_{12}=X_{1}+cX_{2}+\gamma X_{\Psi}$, we have
the invariant solution%
\begin{equation}
\Psi_{12}\left(  A,B\right)  =\exp\left(  -\frac{\gamma}{c}\frac{2B^{4}%
\Lambda+6\omega B^{2}-3\mu}{6B}\right)  \Phi^{\prime}\left(  C^{\prime
}\right)
\end{equation}
in which $\Phi^{\prime}\left(  C^{\prime}\right)  $ satisfies the second-order
differential equation%
\begin{equation}
2\Phi_{,C^{\prime}C^{\prime}}^{\prime}-\gamma\Phi_{,C^{\prime}}^{\prime}%
-c\Phi^{\prime}=0,
\end{equation}
where $C^{\prime}=\frac{3}{\gamma B}\left(  2B^{4}\Lambda+6\omega B^{2}%
-3\mu+3cA^{2}B^{2}\right)  $. Hence we have that
\begin{equation}
\Psi_{12}=\exp\left(  -\frac{\gamma}{c}\frac{2B^{4}\Lambda+6\omega B^{2}-3\mu
}{6B}\right)  \left(  \Psi_{12}^{0}\exp\left(  \gamma_{+}C^{\prime}\right)
+\Psi_{12}^{^{\prime}0}\exp\left(  \gamma_{-}C^{\prime}\right)  \right)  ,
\label{so.01}%
\end{equation}
where $\gamma_{\pm}=\frac{1}{4}\left(  \gamma\pm\sqrt{\gamma^{2}+8c}\right)  $.

Furthermore from the Lie symmetry vector, $Y_{13}=cX_{1}+X_{3}+3mX_{\Psi}%
$,$~m\in%
\mathbb{C}
$,~we find the invariant solution%
\begin{equation}
\Psi_{13}\left(  A,B\right)  =\left(  \Psi_{13}^{0}I_{m}\left(  \frac{\sqrt
{6}}{3}\bar{A}\bar{B}\right)  +\Psi_{13}^{0\prime}K_{m}\left(  \frac{\sqrt{6}%
}{3}\bar{A}\bar{B}\right)  \right)  \left(  \frac{\bar{A}}{\bar{B}}\right)
^{-m}, \label{so.02}%
\end{equation}
where $I_{m},~K_{m}$ are the modified Bessel functions of the first and second
kind respectively.

Furthermore from the symmetry vector $Y_{23}=cX_{2}+X_{3}+3mX_{\Psi}$,~$m\in%
\mathbb{C}
,$ the following invariant solution follows%
\begin{equation}
\Psi_{23}\left(  A,B\right)  =\left(  \frac{AB}{\bar{C}}\right)  ^{m}\left(
\Psi_{23}^{0}I_{m}\left(  \frac{\sqrt{6}}{3}A\bar{C}\right)  +\Psi
_{23}^{0^{\prime}}K_{m}\left(  \frac{\sqrt{6}}{3}A\bar{C}\right)  \right)  ,
\label{so.03}%
\end{equation}
where $\left(  \bar{C}\right)  =\left(  2\Lambda B^{4}+6\omega B^{2}%
+2cB-3\mu\right)  ^{1/2}$.

Finally from the generic symmetry vector, $Y_{G}=c_{1}X_{1}+c_{2}X_{2}%
+X_{3}+3mX_{\Psi}$, we have the following invariant solution%
\begin{equation}
\Psi_{G}\left(  A,B\right)  =\left(  \Psi_{G}^{0}I_{m}\left(  \sqrt
{2}iA^{\prime}B^{\prime}\right)  +\Psi_{G}^{\prime0}K_{m}\left(  \sqrt
{2}iA^{\prime}B^{\prime}\right)  \right)  \left(  \frac{A^{\prime}}{B^{\prime
}}\right)  ^{m_{3}}, \label{so.04}%
\end{equation}
where
\begin{equation}
A^{\prime}=\sqrt{3BA^{2}+2c_{1}}%
\end{equation}
and%
\begin{equation}
B^{\prime}=\frac{1}{3}B^{-\frac{1}{2}}\left(  2\Lambda B^{4}+6\omega^{2}%
B^{2}+2c_{2}B-3\mu\right)  ^{\frac{1}{2}}\text{.}%
\end{equation}
We remark that, as equation (\ref{eq.09}) is a linear equation, the general
solution is expressed as a linear combination of all the Lie invariant
solutions for all the values of the free parameters, that is%
\begin{equation}
\Psi_{Total}\left(  A,B\right)  =%
{\displaystyle\sum\limits_{m_{1}}}
a_{m_{1}}\Psi_{1}\left(  A,B\right)  +%
{\displaystyle\sum\limits_{m_{2}}}
b_{m_{2}}\Psi_{2}\left(  A,B\right)  +... \label{so.05}%
\end{equation}

In the following Section we proceed with the derivation of the analytical
solution for the classical field equations (\ref{eq.1})-(\ref{eq.3}).

\section{Classical Solution}

\label{classicalSolution}

As we discussed in Section \ref{field}, the existence of a Lie point symmetry
for the WdW Equation, (\ref{eq.08}), is equivalent with the existence of a
Noetherian conservation law for the field equation. It is easy to see that the
Lagrangian, (\ref{eq.01}), does not admit any Noether symmetries apart from
the autonomous one. However, for specific lapse, $N$, the conformally related
Lagrangian (\ref{eq.5}) admits as Noether symmetries the Lie point symmetries
of equation (\ref{eq.08}) (for details see \cite{IJGMMP}).

For instance, when $N\left(  A,B\right)  =A^{-1}$,  the vector field, $X_{1},$
is a Noether symmetry of (\ref{eq.5}) and the corresponding conservation law
is $I_{1}=\dot{B}$. \ The Hamiltonian, (\ref{eq.7}), and the conservation law
are in involution, i.e., $\left\{  H,I_{1}\right\}  =0$, which means that the
dynamical system which describes the field equations is Liouville integrable.

Similarly for different lapse, $N$, the other vector fields, i.e. $X_{2}%
,X_{3}$, or any linear combination, produce Noetherian conservation laws. On
the other hand, as the solution is unique and the field equations are
conformally invariant the analytical solution that we find can be transformed
into different lapse functions, $N$.

Under the coordinate transformation,%
\begin{equation}
A=\sqrt{\frac{2x}{y}},~B=y,\label{cl.03}%
\end{equation}
the Lagrangian, (\ref{eq.5}), of the field equations becomes%
\begin{equation}
L=\frac{\sqrt{2}}{2N}\sqrt{\frac{y}{x}}x^{\prime}y^{\prime}+N\sqrt{2}%
\sqrt{\frac{x}{y}}\left(  \omega+\Lambda y^{2}+\frac{\mu}{2y^{2}}\right)
\label{cl.04}%
\end{equation}

and the Hamiltonian
\begin{equation}
H\equiv N\sqrt{2}\sqrt{\frac{x}{y}}\left(  \frac{1}{2}p_{x}p_{y}-\left(
\omega+\Lambda y^{2}+\frac{\mu}{2y^{2}}\right)  \right)  =0. \label{cl.05}%
\end{equation}

Without loss of generality we select the lapse function $N\left(  A,B\right)
=A^{-1}=\sqrt{\frac{y}{2x}}$. \ The line element of the spacetime,
(\ref{eq.0}), becomes
\begin{equation}
ds^{2}=-\frac{1}{A^{2}\left(  \tau\right)  }d\tau^{2}+A^{2}\left(
\tau\right)  dr^{2}+B^{2}\left(  \tau\right)  \left(  d\theta^{2}+\sin
^{2}\theta d\phi^{2}\right)  .\label{cl.06}%
\end{equation}
Moreover, the Hamiltonian (\ref{cl.05}) of the field equations becomes%
\begin{equation}
H\equiv\frac{1}{2}p_{x}p_{y}-\left(  \omega+\Lambda y^{2}+\frac{\mu}{2y^{2}%
}\right)  =0\label{cl.07}%
\end{equation}
and Hamilton's equations are
\begin{equation}
x^{\prime}=\frac{1}{2}p_{y}~,~y^{\prime}=\frac{1}{2}p_{x},\label{cl.08}%
\end{equation}
and%

\begin{equation}
p_{x}^{\prime}=0~,~p_{y}^{\prime}=2\Lambda y-\frac{\mu}{y^{3}}.\label{cl.09}%
\end{equation}
Equations (\ref{cl.07})-(\ref{cl.09}) are the field equations for the model
with the Action, (\ref{eq.00}).

The general solution of the field equations is ($I_{1}\neq0$):
\begin{equation}
y\left(  \tau\right)  =\frac{I_{1}}{2}\tau+y_{0} \label{cl.10}%
\end{equation}%
\begin{equation}
x\left(  \tau\right)  =\frac{I_{1}\Lambda}{12}\tau^{3}+\frac{\Lambda y_{0}}%
{2}\tau^{2}-\frac{2\mu}{I_{1}^{2}\left(  I_{1}\tau+2y_{0}\right)  }+x_{1}%
\tau+x_{0} \label{cl.11}%
\end{equation}
with constraint
\begin{equation}
x_{1}I_{1}-2\Lambda y_{0}^{2}-2\omega=0.
\end{equation}

In the special case for which $I_{1}=0$ the solution is%
\begin{equation}
x\left(  \tau\right)  =\frac{2\Lambda y_{0}^{4}-\mu}{4y_{0}^{3}}\tau^{2}%
+x_{1}\tau+x_{0}~,~y\left(  \tau\right)  =y_{0},\label{cl.13}%
\end{equation}
where now the constraint equation which follows from (\ref{cl.07}) is \
\begin{equation}
\omega+\Lambda y_{0}^{2}+\frac{\mu}{2y_{0}^{2}}=0\label{cl.14a}%
\end{equation}
and so $\omega<0$. The last solution means that $B\left(  t\right)  =y_{0},$
that is, the fluid components, (\ref{TE.3}) and (\ref{TE.4}), of the Skyrme
field are constants.

Hence from (\ref{cl.03}) we have that%
\begin{equation}
A\left(  \tau\right)  =\sqrt{A_{0}\tau^{2}+2\frac{x_{1}}{y_{0}}\tau
+2\frac{x_{0}}{y_{0}}},
\end{equation}
where $A_{0}=\frac{2\Lambda y_{0}^{4}-\mu}{y_{0}^{4}}$, and, when $x_{1}%
=x_{0}=0,$ we have that $A^{2}\left(  \tau\right)  =A_{0}\tau^{2},$ whereas in
the proper time, $N=1$, $A\left(  t\right)  =\sqrt{A_{0}}e^{\sqrt{A_{0}}t}$
which is a de Sitter behaviour. The latter solution holds only when $I_{1}=0$.
\ Hence a small perturbation of the solution gives that $I_{1}\neq0$ and so
the solution is unstable.

Concerning the solution with $I_{1}\neq0$, from (\ref{cl.03}) and
(\ref{cl.10}) we have that%
\begin{equation}
\tau=\frac{2}{I_{1}}B-\frac{y_{0}}{I_{1}}.
\end{equation}
Hence%
\begin{equation}
x\left(  B\right)  =\frac{\Lambda}{12I_{1}^{2}}\left(  2B-y_{0}\right)
^{3}+\frac{\Lambda y_{0}}{2I_{1}^{2}}\left(  2B-y_{0}\right)  ^{2}-\frac{\mu
}{I_{1}^{2}B}+\frac{x_{1}}{I_{1}}\left(  2B-y_{0}\right)  +x_{0}%
\end{equation}
which means that $A\left(  \tau\right)  $ is expressed as a function of $B$,
i.e. $A\left(  B\right)  $.

\begin{figure}[ptb]
\includegraphics[height=7cm]{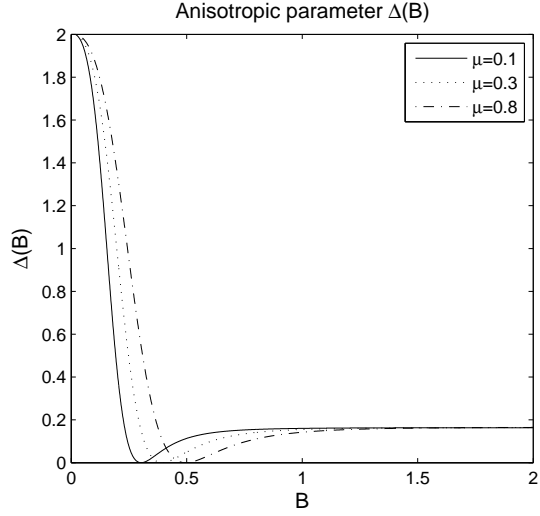}
\caption{Evolution of the anisotropic parameter~$\Delta\left(  B\right)  $ for
various values of the parameter $\mu$. We observe that the anisotropic
parameter reach a minimum close to zero and after it becomes constant. For the
plots we selected $A=0$ when $B=0$.}%
\label{fig3}%
\end{figure}

For the line element, (\ref{cl.06}), we can define the spatial volume
$V=AB^{2}\sin\theta$ and the average scale factor $a=V^{\frac{1}{3}}%
~$\cite{Yadav}. Therefore the average Hubble function, $H=A\frac{a^{\prime}%
}{a},$ is%
\begin{equation}
H=\frac{1}{3}\left(  A^{\prime}+2\frac{A}{B}B^{\prime}\right)  ,\label{cl.15}%
\end{equation}
where $H_{A}=A^{\prime},~H_{B}=\frac{A}{B}B^{\prime}$ are the Hubble functions
on the direction $r$ and on the two sphere $d\Omega^{2},$ respectively.
However,
\begin{equation}
A^{\prime}=\frac{1}{A}\left(  \frac{x^{\prime}}{y}-\frac{x}{y^{2}}y^{\prime
}\right)  =\left(  \frac{x^{\prime}}{AB}-\frac{I_{1}}{4}\frac{A}{B}\right)
\label{cl.16}%
\end{equation}
and $B^{\prime}=\frac{I_{1}}{2}$. Therefore equation (\ref{cl.15}) becomes%
\begin{equation}
H=\frac{1}{3}\left(  \frac{x^{\prime}}{AB}+\frac{I_{1}}{4}\frac{A}{B}\right)
,\label{cl.17}%
\end{equation}
where%
\begin{equation}
\frac{dx\left(  B\right)  }{dB}=\frac{\Lambda}{4I_{1}}\left(  2B-y_{0}\right)
^{2}+\frac{\Lambda y_{0}}{I_{1}}\left(  2B-y_{0}\right)  +\frac{2\mu}%
{4I_{1}^{2}B^{2}}+x_{1}.\label{cl.18}%
\end{equation}
Finally the anisotropic parameter is defined as
\begin{equation}
\Delta\left(  B\right)  =\frac{\left(  H_{A}-H\right)  ^{2}+2\left(
H_{B}-H\right)  ^{2}}{3H^{2}}.\label{cl.19}%
\end{equation}
The evolution of the anisotropic parameter, $\Delta\left(  B\right)  ,$ can be
found in fig. \ref{fig3}, where it is shown that the anisotropic parameter
vanishes at a point and afterwards becomes a nonzero constant.

\subsection{Semiclassical Solution}

The solution presented above is the classical solution of the field equations.
Because the solution of the WdW equation is known we study the quantum effects
in the classical behaviour which follow from the quantum potential \ in the
semiclassical approach of Bohmian mechanics \cite{boh1,boh2}.  A method which
has been applied recently in various models \cite{dimakisT2,TchRN,manto,Stu2}.

In particular, if the solution of the WDW equation is $\Psi\left(
x^{k}\right)  =A\left(  x^{k}\right)  e^{iB\left(  x^{k}\right)  }$, where now
$A\left(  x^{k}\right)  $ is not a slow-roll function, then the substitution
of that solution into the WDW equation gives
\begin{equation}
H\left(  p_{k},x^{k}\right)  +Q_{V}=0,\label{boh.1}%
\end{equation}
where $H\left(  p_{k},x^{k}\right)  $ is the Hamiltonian function which
generates the WDW equation, $Q_{V}=\frac{\Delta\left(  A\right)  }{A},$ is the
quantum potential and the momentum $p_{\mu}=\frac{\partial B\left(
x^{k}\right)  }{\partial x^{\mu}}$. \ Equation (\ref{boh.1}) is the new
Hamiltonian of the field equations which provides us with the semiclassical
solution. \ In the limit for which $Q_{V}=0$ or $A$ is a slow-roll function we
are in the region of the WKB approximation if and only if $B\left(
x^{k}\right)  $ is a solution of the Hamilton-Jacobi equation for the
classical field equations.

From the solution of the WdW Equation, (\ref{so.01}), we can see that the
quantum potential is zero and we are in the classical solution. Consider now
the wavefunction (\ref{so.03}) and $\Psi_{23}^{0\prime}=0$. The wavefunction
is
\begin{equation}
\Psi\left(  A,B\right)  =\Psi_{23}^{0}\left(  \frac{AB}{\bar{C}}\right)
^{m}I_{m}\left(  \frac{\sqrt{6}}{3}A\bar{C}\right)  .\label{so.m1}%
\end{equation}

In order to determine the quantum correction we have to extract from the
wavefunction the nonoscillatory terms. \ Let not $m=i\sigma$, $\sigma\in%
\mathbb{R}
$, and we are in the region in which~$A\rightarrow0$,~$B\rightarrow0,~\ $such
that $B<<\mu,$ which gives that $\bar{C}=i\left\vert \bar{C}\right\vert
.~$Hence the solution (\ref{so.m1}) is approximated as follows%
\begin{equation}
\Psi\left(  A,B\right)  \simeq\bar{\Psi}_{0}\exp\left[  i\sigma\ln\left(
AB\right)  \right]  \exp\left(  -i\bar{\sigma}\ln\left(  \left\vert \bar
{C}\right\vert \right)  \right)  \left(  A\left\vert \bar{C}\right\vert
\right)  ^{\sigma},\label{so.m2}%
\end{equation}
which provides us with the quantum potential~$Q_{V}=\frac{3\sigma^{2}\left(
2\Lambda B^{4}+2\omega B^{2}+\mu\right)  }{4AB^{2}}\left\vert \bar
{C}\right\vert ^{-2}$. \ Here we note that, if $\bar{C}$ was a real function,
then the quantum potential would be zero. \ Furthermore from (\ref{so.m2}) we
have the oscillatory term, $B\left(  A,B\right)  =\sigma\ln\left(  AB\right)
-\bar{\sigma}\ln\left(  \left\vert \bar{C}\right\vert \right)  ,$ from which
we find that $p_{A}=\frac{\sigma}{A}~,~p_{B}=\frac{\sigma}{B}+\frac
{\bar{\sigma}\left(  8\Lambda B^{3}+12\omega B-2c\right)  }{2\bar{C}^{2}}$ and
the reduced field equations are
\begin{equation}
\frac{2}{N}BB^{\prime}=\frac{\sigma}{A},\label{so.m3}%
\end{equation}%
\begin{equation}
\frac{2}{N}BA^{\prime}=\frac{\bar{\sigma}\left(  8\Lambda B^{3}+12\omega
B-2c\right)  }{2\bar{C}^{2}}.\label{so.m4}%
\end{equation}

If we consider that $N=A$, we have the solution%
\begin{equation}
B^{2}\left(  t\right)  =2\sigma\left(  \tau-\tau_{0}\right)
\end{equation}
and%
\begin{equation}
\ln\left(  A\left(  t\right)  \right)  =\int\frac{\bar{\sigma}\left(  8\Lambda
B^{3}+12\omega B-2c\right)  }{2B\bar{C}^{2}}%
\end{equation}
which is different from the previous solution. Of course that solution holds
only in the region of $A,B$ which we have already considered for the
derivation of the quantum potential and specifically when $A\rightarrow
0$,~$B\rightarrow0,$ that is, quantum effects take place\textbf{.}

\section{Conclusions}

\label{Conc}

In this work we considered a locally rotational Kantowski-Sachs spacetime with
cosmological constant and a Skyrme fluid with constant radial profile. For the
comoving observer the Skyrme fluid is an anisotropic fluid with zero heat flux
component and nonzero stress tensor. Furthermore for the equation-of-state
parameter, $w_{S},$ for the Skyrme fluid $\left\vert w_{s}\right\vert
\leq\frac{1}{3}$  holds.

In order to construct the solution of the field equations we followed a method
established in \cite{AB}, that is, we derived the WdW Equation and we showed
that it admits Lie point symmetries. The existence of the Lie point symmetries
means that the wavefunction of the universe admits oscillatory terms which
follow from the Lie invariants. Moreover, from the Lie point symmetries of the
WdW Equation we constructed Noetherian conservation laws for the classical
field equations and proved the Liouville integrability of the field equations
We used the Noetherian conservation laws in order to derive the solution of
the field equations. We have found two families of solutions, one in which the
scale factor, $B$, of the two sphere, is not constant and one in which it is
constant. \ For the first solution we expressed the cosmological parameters in
terms of the scale factor, $B$, and we derived the anisotropic parameter,
$\Delta\left(  B\right)  ,$ and showed that for initial conditions
$\lim\limits_{B\rightarrow0}A\left(  B\right)  =0$ the anisotropic parameter
reaches a minimum close to zero and for large values of $B$ becomes constant
different than zero. For the second solution for which $B=const.$ we have
shown that the scale factor, $A$, of the radius, has an exponential behaviour.

At this point we wish to compare our results with the solutions that have been
found previously in \cite{Parisi}. As we discussed above, in the latter work
the authors performed a fixed point analysis of the gravitational field
equations, (\ref{eq.1})-(\ref{eq.3}), and they derived some special solutions
which describe the solution of the field equations at the fixed point, i.e.
for a

specific initial-value problem of the field equations. In this work the
solutions that we have presented are the general solution of the system and
that can be seen from the number of the constants of integration. If we
consider the initial conditions to be that of the fixed points of the field
equations, the special solution of \cite{Parisi} is easily recovered.
Furthermore the importance of this work is that the integrability of the field
equations (\ref{eq.1})-(\ref{eq.3}) is proved  which means that a solution
always exists, while integrability cannot proved by the fixed-point analysis.
Only in the region of the fixed points can the solution be approximated from
the linearized system. The integrability of the Einstein-nonlinear $\sigma$
models is still an open subject of special interest.

Before we conclude, we note that the WdW Equation (\ref{eq.09}) follows from a
variational principle in which the Lagrangian for the WdW Equation is
\begin{equation}
\mathcal{L}\left(  A,B,\Psi_{,A},\Psi_{,B}\right)  =-\frac{A}{4B^{2}}\left(
\Psi_{,A}\right)  ^{2}+\frac{1}{2B}\Psi_{,A}\Psi_{,B}+\frac{1}{2}\left(
AB^{2}\Lambda+\omega A+\mu\frac{A}{2B^{2}}\right)  \Psi^{2}.\label{cl.20}%
\end{equation}

In \cite{IJGMMP2} it has been shown that the nontrivial Lie point symmetries
of the Klein-Gordon equation are also Noether symmetries. Hence from the Lie
point symmetries, (\ref{eq.13})-(\ref{eq.15}), and for the Lagrangian
(\ref{cl.20}) we can construct conservation flows for the wavefunction of the
universe. In particular, for the case in which the minisuperspace has
dimension two, the Noetherian conservation flow is given by the expression
$I^{i}=\xi^{k}H_{k}^{i}$ $,$ and satisfies the condition $D_{i}I^{i}=0$,
where
\begin{equation}
\mathcal{H}_{k}^{i}=\frac{1}{2}\left(  \frac{\partial\mathcal{L}}{\partial
\Psi_{,i}}\Psi_{,k}-\delta_{k}^{i}\mathcal{L}\right)  ,
\end{equation}
and, $D_{i}=\frac{\partial}{\partial x^{i}}+\Psi_{,i}\frac{\partial}%
{\partial\Psi}+\Psi_{,ij}\frac{\partial}{\partial\Psi_{j}}+$...$.$

For the Lagrangian (\ref{cl.20}) we have that $x^{i}=\left(  A,B\right)  $.
Hence the conservation flow which corresponds to the symmetry vector
$X_{1},~X_{2},~X_{3}$ has the following components
\begin{equation}
\mathcal{I}^{A}\left(  X_{1}\right)  =-\frac{1}{2}\frac{\left(  \Psi
_{,A}\right)  ^{2}}{B^{3}}+\frac{V_{eff}}{AB}\Psi^{2}~,~\mathcal{I}^{B}\left(
X_{1}\right)  =\frac{\left(  \Psi_{,A}\right)  ^{2}}{AB^{2}},
\end{equation}%
\begin{equation}
\mathcal{I}^{A}\left(  X_{2}\right)  =\frac{\left(  A\Psi_{,A}-2B\Psi
_{,B}\right)  ^{2}-2AB^{2}V_{eff}\Psi^{2}}{\left(  2\Lambda B^{4}+2\omega
B^{2}+\mu\right)  B}~,~
\end{equation}%
\begin{equation}
\mathcal{I}^{B}\left(  X_{2}\right)  =\frac{4B^{2}V_{eff}\Psi^{2}}{\left(
2\Lambda B^{4}+2\omega B^{2}+\mu\right)  }%
\end{equation}
and%
\begin{equation}
\mathcal{I}^{A}\left(  X_{3}\right)  =-\frac{\left(  2\Lambda B^{4}+2\omega
B^{2}+\mu\right)  ^{-1}}{2}\left[
\begin{array}
[c]{c}%
A\left(  2\Lambda B^{2}+3\omega\right)  \left(  A\left(  \Psi_{,A}\right)
^{2}-B^{2}V_{eff}\Psi^{2}\right)  +\\
+\left(  2\Lambda B^{4}+6\omega B^{2}-3\mu\right)  \left(  \left(  \Psi
_{,B}\right)  ^{2}-AB^{-1}\Psi_{,A}\Psi_{,B}\right)
\end{array}
\right]
\end{equation}%
\begin{equation}
\mathcal{I}^{B}\left(  X_{3}\right)  =\frac{3}{4}\frac{A}{B}\left(  \Psi
_{,A}\right)  ^{2}-\frac{B}{2}\frac{\left(  2\Lambda B^{4}+6\omega B^{2}%
-3\mu\right)  }{\left(  2\Lambda B^{4}+2\omega B^{2}+\mu\right)  }V_{eff}%
\Psi^{2}%
\end{equation}
where $V_{eff}=\left(  AB^{2}\Lambda+\omega A+\mu\frac{A}{2B^{2}}\right)  .$

Therefore, the existence of Lie point symmetries for the WdW equation is
equivalent with the existence of Noetherian conservation laws for the latter equation.

A more general consideration of the Skyrme fluid with nonconstant radial
profile \cite{Canfora4} will be of interest. Such an analysis is in progress
and will be published in a forthcoming work.

\begin{acknowledgments}
The research of AP was supported by FONDECYT grant no. 3160121.
\end{acknowledgments}

\end{document}